\documentclass[aps,prl,twocolumn,showpacs,superscriptaddress,floatfix,preprintnumbers]{revtex4-1}
\usepackage{epsfig,amssymb,amsmath,amsfonts}
\usepackage{graphicx}
\usepackage{color}


\def\l{\lambda}

\def\1{\bf{1}}

\def\th{\theta}

\def\erfc{\text{erfc}}

\begin{document}
\title{Large deviations of a tracer in the  symmetric exclusion process} 
\author{Takashi Imamura}
\affiliation{Department of Mathematics and Informatics,
 Chiba University, 1-33 Yayoi-cho, Inage, Chiba 263-8522, Japan}
\author{Kirone Mallick}
\affiliation{Institut de Physique Th\'eorique,  CEA Saclay
and URA 2306, CNRS, 91191 Gif-sur-Yvette cedex, France}  
 \author{Tomohiro Sasamoto}
\affiliation{Department of Physics, Tokyo Institute of Technology, 2-12-1
 Ookayama, Meguro-ku, Tokyo 152-8551, Japan}

\date{\today}
\begin{abstract}
 The one-dimensional symmetric exclusion process, the simplest
 interacting particle process,  is  a lattice-gas made  of  particles
 that hop  symmetrically  on a discrete line   respecting
 hard-core exclusion. The system  is prepared on the infinite lattice with  a  step  initial
 profile with  average densities $\rho_{+}$ and  $\rho_{-}$ on the
 right and on the left  of the origin. When  $\rho_{+} = \rho_{-}$,
 the gas is at equilibrium and undergoes
 stationary fluctuations.  When   these  densities are
 unequal, the gas is out of equilibrium and will remain so forever.  A
tracer, or a tagged particle, is initially  located at the boundary
 between the two domains; its  position $X_t$  is a random  observable
 in  time, that carries information  on the non-equilibrium dynamics
 of the whole system.  We derive   an  exact
 formula for the cumulant generating function and the large deviation
 function of $X_t$, in  the long time limit, 
  and deduce  the  full statistical properties of
 the tracer's position. The equilibrium fluctuations of the
 tracer's position,  when the density is uniform, 
  are obtained as an important special case. 
 \end{abstract}
\maketitle

 The collective dynamics of a  complex system   can be probed by
 attaching  a  neutral tag to a  particle, that does not  alter its
 interactions with the   environment,  and by    monitoring the
 position  of the tagged particle in   time.  This  technique  is a
 powerful tool to study flows in  material sciences,  biological
 systems and even social groups (see e.g.,
 \cite{ChaikinBook,KARGER1992,Tsien98,SchadBook} and references
 therein).  The averaged  trajectory  of a  tracer carries
 information  on the  overall  motion of the fluid whereas its
 fluctuations are sensitive to the statistical properties of the
 medium. The canonical example is  the  Brownian motion of a grain of
 pollen immersed in  water  at thermal equilibrium,   and the
 simplest  model for this  diffusion  is given by  independent random
 walkers  symmetrically hopping on a lattice;  the position of any
 walker, as a function of time $t$,   spreads as $\sqrt{t}$. In   presence of
 weak-interactions, diffusive behavior   generically prevails but
 the amplitude of the spreading, measured by   the  diffusion
 constant, is  a function of   the total density of particles
 \cite{SpohnBook,KrapivskyBook,ChaikinBook}.

 If the interactions induce long-range correlations either in space or time direction, 
 or if the environment  is  out  of equilibrium (by  carrying   some internal
 currents), the motion of  a   tagged  particle can  exhibit unusual
 statistical properties such as anomalous diffusion and/or
 non-Gaussian fluctuations.  For example, a  tracer trapped in a
 linear array of  convection rolls   spreads only  as $t^{1/3}$  with
 time \cite{YoungPumirPomeau1989,BouchaudGeorges1990}.  Correlations
 are usually enhanced  in  low dimensional systems such as   narrow
 quasi-one-dimensional channels,  in which  the order amongst the
 particles is preserved because of steric hindrance.  For such a
 single-file motion,  the typical displacement  $X_t$ of a tracer at
 large times grows  as  $t^{1/4}$ , which is much slower than  the
 usual $\sqrt{t}$ law, regardless of  the precise form of the
 interaction.  However, collective diffusion of  local density
 fluctuations remains normal  and  behaves  as $\sqrt{t}$.  
 Similarly,  the time-integrated current at a given location of a 
 single-file channel also displays  $t^{1/4}$-fluctuations.  This
 anomalous  {\it single-file  diffusion}   has been demonstrated  in
 various  experiments involving  different types of 
 physical systems such as zeolites, capillary pores, carbon nanotubes
 or colloids
 \cite{HODGKIN1955,Lea1963,KUKLA1996,CHOU1999,WBL2000,Kollmann2003,Lin2005}.
 Single-file  diffusion is also  discussed  in  numerous  theoretical
 papers, at various levels of  physical intuition
 \cite{Levitt1973,*Percus1974,*Richards1977,*Alexander1978,*VanBeijeren1983,*Majumdar1991,
*Rodenbeck1998,*Leibovich2013,*Illien2013,*Lizana2014}
 or mathematical rigor
\cite{Harris1965,*Spitzer1970,*Arratia1983,*Spohn1990,SpohnBook,Liggett2004}.

 One of the  simplest models in  non-equilibrium statistical physics
 is   the Symmetric Exclusion Process (SEP) \cite{Spitzer1970},  a
 lattice gas  of particles  performing   symmetric
 random walks  in  continuous time 
  and  interacting  by hard-core exclusion: each particle
 attempts to hop  with rate unity from its location  to an empty
 neighboring site;  double occupancy of a site is forbidden. Thanks to
 the  wealth of  analytical knowledge   accumulated  during the last
 few decades,  this  process  and its variants,   are  used as
  paradigms  in   non-equilibrium statistical mechanics 
 \cite{KLS1984,Presutti88,Derrida2007,Derrida2011,KrapivskyBook}.  In
 a one dimensional lattice, the  SEP is  a pristine model of a
 single-file diffusion, amenable to quantitative analysis. 
 In  equilibrium  case with uniform density  $\rho$,  the variance of the
 position  $X_t$ of a  tagged particle initially located  at  the
 origin   is given, in the long time limit,
 by   \cite{Arratia1983,SpohnBook}
\begin{equation}
 \langle X_t^2\rangle = \frac{2(1-\rho)}{\rho}\sqrt{\frac{t}{\pi}} . 
 \label{var}
\end{equation}
  It has also  been proved 
 that  the {\it rescaled}  position  $\frac{X_t}{t^{1/4}}$ satisfies a central limit theorem 
 and converges to a fractional Brownian motion  with Hurst index 1/4
 \cite{Arratia1983, SethuramPeligrad}. 

 The full distribution of $X_t$  and its higher cumulants are,
 however, not known. The tracer, being  immersed  in   fluctuating
 environment, far from equilibrium, can display large and non-typical
 excursions.  Such  rare events are quantified by  a large deviation
 function \cite{Touchette,DemboZeitouni2009}.  Large deviation functions appear as
 appropriate  candidates for macroscopic potentials under  non-equilibrium
 conditions. Moreover,  the fluctuation theorem, which 
 is one of the few exact results
  valid  far  from thermodynamic equilibrium, is stated 
 as a property of   large deviation functions 
 \cite{GallavottiCohen1995,LebowitzSpohn1999}. It   emphasizes    the role
 of microscopic  time-reversal symmetry  for macroscopic fluctuations.
  In present
 day  statistical physics,  large deviations  play   an increasingly
 important role \cite{Derrida2007,Derrida2011,Jona-Lasinio2014,Bertini2014}.

 Recently,  the  large deviation principle  for the tracer position
 has been proved rigorously   \cite{Sethuraman2013}: when $t \to
 \infty$ there exists a  large-deviation function  $\phi(\xi)$, such that 
\begin{equation}
 \text{Prob}\left( \frac{X_t}{\sqrt{4 t}} = -\xi  \right)
  \sim  \exp[-\sqrt{t} \phi(\xi)].
\label{def:phi}
\end{equation} 
 Note the prefactor $\sqrt{t}$  in the exponent; for non-interacting
particles, the  prefactor would be  $t$    \cite{Supplementary}.
Alternatively, one  studies   the characteristic  function of $X_t$,
 which behaves as 
\begin{align}
\langle e^{s X_t} \rangle \sim e^{- \sqrt{t} C(s)} \quad {\rm when } \quad  t \to \infty.
\end{align}
 The  Taylor expansion of  the cumulant generating function
 $C(s)$ with respect to  $s$  generates  all the cumulants
 of $X_t$.    The functions   $C(s)$ and   $\phi(\xi)$  are
 related by  Legendre transform \cite{Touchette,DemboZeitouni2009}:
 \begin{align}
  C(s) = \min_{\xi} \left( 2s \xi  + \phi(\xi)  \right).
  \label{Cs}
\end{align}
Each of these functions  carries  information on the
long time behavior of the  process.  Although the SEP has been
studied for more than 40 years, analytic formulas for these  functions
are not  yet known.
  
 In this letter,  we  shall present an exact formula for the
large-deviation function $\phi(\xi)$ in (\ref{def:phi}) 
 of  the tracer position in the SEP.
As an initial condition, we prepare a step density profile
with  an average density  $\rho_{+}$  on the right of the origin and 
   $\rho_{-}$  on the left. (See the right figure in Fig. 2.)
In a parametric representation, $\phi(\xi)$ is given by 
\begin{align}
\begin{cases}
  \phi(\xi)  & =  \mu(\xi, \lambda^*),    \\
 \frac{ \partial \mu(\xi, \lambda^*)} {\partial  \lambda}  & =  0,
\label{def:lambda}
\end{cases}
\end{align} 
where the second equation defines implicitly  $\lambda^* =\lambda^*(\xi)$, 
and $\mu(\xi,\l)$ is 
\begin{align} 
& \mu(\xi,\l) =
 \sum_{n=1}^{\infty}  \frac{ (-\omega)^n}{n^{3/2}}  A(\sqrt{n}\, \xi) \,
 +  \xi \log\frac{ 1+\rho_+(e^{\lambda}-1)} 
{1+\rho_-(e^{-\lambda}-1)}.
\label{cumgen}
\end{align}
Here 
$\omega   =   r_+(e^\l-1)+r_-(e^{-\l}-1)$ with  
$r_{\pm} = \rho_{\pm}(1-\rho_{\mp})$
and
\begin{align}
    A(\xi)   
&=\frac{e^{-\xi^2}}{\sqrt{\pi}}+\xi(1-\erfc{(\xi)}),
\end{align}
where the complementary error function is defined by 
$\erfc{(z)} = \frac{2}{\sqrt{\pi}} \int_z^\infty e^{-u^2}  {\rm d}u$.  
This is the central result in this letter. 
For  $\xi=0$, we have  $A(0)= 1/\sqrt{\pi}$ and $\mu(0,\l)$
 reduces  to the expression found in 
 \cite{Gerschenfeld2009Bethe,*Gerschenfeld2009} for the 
 current fluctuations in the SEP at the origin. 
Our formula (\ref{cumgen}) generalizes it and leads us to a complete analytic description 
of the statistical properties of the tracer in the long time limit. The figure is also 
easily drawn, see Fig 1. 

 \begin{figure}[t]
 \includegraphics[scale=0.5]{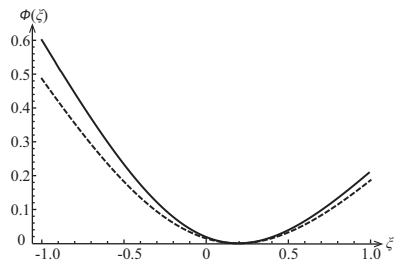}
 \caption{\label{phis}The large deviation function $\phi(\xi)$ of the tracer position 
 in the SEP (solid curve) for the case $\rho_+=0.3, \rho_-=0.15$. 
 The one for the reflective Brownian particles~\eqref{ldrBM} 
with the same $\rho_{\pm}$
 is also shown (dashed curve).    
  }
 \end{figure}

 To explain the meaning of $\mu$ and the derivation of our formula, 
  we  first recall the set-up of  the asymmetric simple
  exclusion process (ASEP), see Fig. 2.  The position of a
  particle is  labeled by an integer  $x \in  \mathbb{Z}$.  Particles
   hop to  the right  and  to the left  with 
  rates  $p$  and  $q$, respectively.
  The asymmetry parameter is
  $\tau=p/q$ with $0 \le \tau \le 1$. The
  symmetric model, which is the main target of our study,  
   corresponds to   $p = q = \tau =  1$.  We adopt  the  convention that a 
  current flowing from right to left is counted 
    positively   \cite{TW2008a, BCS2014}.
The  initial condition    is  the step density profile with $\rho_+$ and $\rho_-$. 
  Typically, we have $\rho_{+} \ge  \rho_{-}$.
  The  stationary case corresponds to   $\rho_{+} = \rho_{-} = \rho$. 
   We  emphasize that  the initial profile  displays randomness:
  statistical averages  will be taken both  over the dynamics  and
  the initial conditions.
 \begin{figure}[t]
 \includegraphics[scale=0.5]{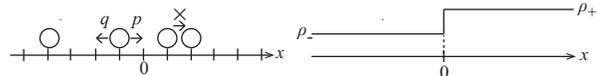}
 \caption{ASEP. Left: Particles hop asymmetrically on the lattice under volume exclusion. 
  Right: Step initial condition with densities $\rho_+$ (resp. $\rho_-$) to the right (resp. left). }
 \end{figure}
The tracer is defined to be the particle in the region $x >0$, which is initially  
the  closest to  the origin.
 (For quantities in the long time, which we are interested in this article, 
this is equivalent to putting the tracer at the origin at $t=0$. )
Its  position at time $t$ is denoted by  $X_t$. 

 In order to study   the   position  of  the continuously moving tracer, 
 it is  useful  to relate  $X_t$ to a local
 observable.  Let $N_t$ denote the integrated current through the bond
 $(0,1)$  for the  duration $[0,t]$; i.e., 
 $N_t$ is equal to the total number of
 particles having hopped from 1 to 0  minus  the total number of
 particles having hopped from 0 to 1  during  the time interval  $[0,t]$.
 We   define  the  following  quantity \cite{Schuetz1997a,ImamuraDuality}
 \begin{equation}
 N(x,t) = N_t+ \begin{cases}
                             + \sum_{y=1}^x \eta_y(t) \,, & x>0, \\
                              0, & x=0, \\
                             -  \sum_{y=x+1}^0 \eta_y(t) \,, & x<0 ,       
                        \end{cases}
\label{def:N}
\end{equation}
 where $\eta_x(t)=1$(resp. 0) when the site $x$ is occupied (resp. empty)
 at time $t$. We recall that the  observable  $h(x,t)= N(x,t)-  x/2$ 
 is   the local height function   appearing  when  the ASEP is mapped  to  
 a growth process \cite{KrapivskyBook,HalpinHealy1995,PS2002a}.

 Using particle  number conservation, one can verify \cite{Supplementary}
 that the  tagged particle position $X_t$ and  $N(x,t)$ satisfy 
\begin{equation}
\mathbb{P}[X_t \leq x] = \mathbb{P} [N(x,t)  >  0] \,.
\label{XversusN}
 \end{equation} 
 This  identity  will  allow
  us to relate the statistical properties  of  the tracer $X_t$ with  those  of 
 the height  $N(x,t)$ and,   in  particular, to express 
 the  cumulant generating function and  the  large deviation   function
  of  $X_t$  in terms of  the corresponding quantities  for $N(x,t)$. 
 
In the long time limit, 
the  characteristic function of $N(x,t)$ behaves as 
\begin{equation}
 \langle e^{\lambda N(x,t)} \rangle \sim e^{- \sqrt{t} \mu(\xi,\l)}  
\label{def:xi}
\end{equation}
with  $\xi = - \frac{x}{\sqrt{4 t}}$
and  its  cumulants  are obtained by  expanding    $\mu(\xi,\l)$ with respect to $\l$.   
This $\mu(\xi,\l)$ is nothing but the one in (\ref{cumgen}). 
From  the identity \eqref{XversusN}, we see  \cite{Supplementary} that 
the large deviation functions of $X_t$ is given through the characteristic function of $N(x,t)$ as 
$ \phi(\xi) 
             = \max_{\l} \mu(\xi,\l)$,  
which is equivalent to (\ref{def:lambda}).

 We now investigate some  properties of the above formulas  and
 extract some concrete results from them. We also  
 retrieve  and generalize  some  results previously known
 in certain  particular cases.

 The  tracer's large deviation function $\phi(\xi)$
 satisfies a version of the Fluctuation Theorem 
  \cite{GallavottiCohen1995,LebowitzSpohn1999},
\begin{equation}
 \phi(\xi)-\phi(-\xi)= 2 \xi \log\frac{ 1 - \rho_{+}}{1 - \rho_{-}}.
\label{FT}
\end{equation}
 The   Fluctuation Theorem  is a symmetry relation
 that originates  from  an underlying time-reversal invariance.
 It  implies,
 in particular, that the Einstein relation is true for the SEP
   \cite{Ferrari1985}.  The proof  of \eqref{FT}
  is based   on  the fact that 
  $\lambda^*(-\xi)  = \log \frac{r_{-}}{r_{+}}\ - \lambda^*(\xi)$ 
 \cite{Supplementary}. 
We also note that, while the fluctuation theorems have been established 
mainly for a large system in the infinitely late time,  ours  
is for a system on the infinite lattice and for a large time. 
 
 Explicit formulae for the first few  cumulants of  $ X_t$
 can be obtained by substituting an expression of $\phi(\xi)$ in 
 (\ref{def:lambda}) into (\ref{Cs}). 
  For a stationary initial condition,  $\rho_{+} =\rho_{-} = \rho $,
we have calculated  
 the first few cumulants: 
 the  variance is given by \eqref{var} and  at the  fourth order, we find 
\begin{equation}
 \frac{\langle X_t^4 \rangle_c}{\sqrt{4 t}} = 
  \frac{1-\rho}{\sqrt{\pi} \rho^3}
   [1-(4-(8-3\sqrt{2}) \rho)
  (1-  \rho)+\dfrac{12}{\pi}(1-\rho)^2] \nonumber 
\end{equation}
(the  subscript $c$ indicates a cumulant), in agreement 
 with   calculations 
 based on the Macroscopic Fluctuation Theory (MFT) \cite{KMS2014}.
 Considering the MFT is a description at the level of hydrodynamics, 
this coincidence provides a highly nontrivial check of the MFT.
The procedure  can be carried out  to higher  orders in $s$
   \cite{Supplementary}. 

 For non-equilibrium initial conditions,   $\rho_{+}  >\rho_{-} >0$, 
 the   tracer drifts away from the origin as 
\begin{equation}
 \frac{\langle X_t \rangle}{\sqrt{4 t}} = - \xi_0, 
\end{equation} 
where $\xi_0$ is the unique solution of 
\begin{equation}
  2\xi_0  \rho_{-}  = 
 (\rho_{+}  - \rho_{-})  \int_{\xi_0}^\infty \erfc{(u)}  {\rm d}u.
\label{xi0r}
\end{equation} 
 Solving  \eqref{def:lambda} around $\xi_0$  leads to  the variance
 of the tracer
\begin{equation}
 \mathrm{Var} (X_t) =  \frac{ 4 K (\rho_{+}  - \rho_{-})^2  A(\xi_0) \sqrt{ t}}
{(\rho_{+} \erfc{(\xi_0)} + \rho_{-} \erfc{(-\xi_0)})^2}
\nonumber
\end{equation} 
with 
$$K =
 \frac{\rho_{+}^3 +  \rho_{-}^3 - 3 \rho_{+}^2 \rho_{-} -3 \rho_{+} \rho_{-}^2
 + 4 \rho_{+} \rho_{-}}  { (\rho_{+} +\rho_{-}) (\rho_{+}  - \rho_{-})^2}   
-  \frac{A(\sqrt{2} \, \xi_0)}{\sqrt{2} A(\xi_0)}.
$$

 In the special case $\rho_{-} = 0$, the tracer is the left-most particle of a 
  SEP expanding  in a half-empty space and finding 
 the  distribution of $X_t$  becomes identical to  a  problem
 in   extreme value statistics. From  
 the above expressions, it can be shown that 
  $\langle X_t \rangle \sim \sqrt{ t \log t} $ and 
   $ \mathrm{Var} (X_t) \sim \frac{t}{\log t}$. The tracer 
 follows  a Gumbel law, which is well-known to appear for independent 
 walkers, in spite of interaction effects in the SEP \cite{Arratia1983,Sanjib2007}. 
 
 In  the low density limit $\rho_{-}, \rho_{+} \ll 1$, the SEP
 becomes equivalent to an ensemble of reflecting  Brownian particles
  \cite{Harris1965}.  This system can be viewed as independent
  Brownian motions that 
 exchange their labels when they collide and has been solved  exactly 
  using various techniques. Retaining
 only the first order  terms in $\rho_{\pm}$ in the formula  \eqref{cumgen}, and
  using (\ref{def:lambda}), we obtain the  large deviation of  a tracer
 in the reflecting Brownian limit:
 \begin{equation}
 \phi(\xi) =\left\{ \sqrt{ {\rho_+}   \Xi(\xi) } -
  \sqrt{{\rho_-} \Xi(-\xi)} \right\}^2
\label{ldrBM}
\end{equation}
where  $\Xi(\xi) =  \int_{\xi}^\infty \erfc{(u)} {\rm d}u$.
  This generalizes  the known  result  in  the uniform case  $\rho_+  =  \rho_-$
 \cite{Rodenbeck1998,KMS2014,Hegde2014,DerridaSadhu}. 
A figure of this large deviation function is also drawn in~Fig.~\ref{phis}.
By comparing to the one for the SEP, the effect of interaction among particles
of the SEP is clearly seen.

 In the last part of this work, we outline  the derivation of the 
 main  formula  \eqref{cumgen}.
  The strategy is to find  exact   expressions  for all the  moments  of 
  $N(x,t)$ and then  construct  the cumulant generating function $\mu(\xi,\l)$. 
  The  time evolution equations for the moments 
  of $N(x,t)$ form   a hierarchy of
 coupled  differential equations that must be solved simultaneously.
 This seems to be a daunting task.
  
  Our strategy is to make a detour though the ASEP, with   $\tau <1$, 
  for which the  observable   $N_{\tau}(x,t)$,
  defined  in \eqref{def:N},  satisfies a remarkable {\it self-duality} property
  \cite{Schuetz1997a,ImamuraDuality,BCS2014}. 
 For $x_1 < x_2 < \ldots < x_n$,   $n$-point correlations  of the type 
 $$ \phi(x_1,\ldots,  x_n; t)=  
  \langle \tau^{N_{\tau}(x_1,t)} \ldots  \tau^{N_{\tau}(x_n,t)} \rangle $$
 follow the same   dynamical equations 
 as  the ASEP with a finite number $n$ of particles located at 
  $x_1,  \ldots,  x_n$.  Using the fact that the  ASEP with $n$ particles
  is  solvable  by Bethe Ansatz, 
 these  $\tau$-correlations  can be expressed   as a multiple contour
 integral in the complex plane \cite{TW2008a,BC2014,BCS2014}.
 For the step initial condition with the densities $\rho_{\pm}$, we can write 
\begin{align}
&\langle \tau^{n N_\tau(x,t)} \rangle
  =  \tau^{-n \frac{x}{2}}\tau^{n(n-1)/2}
  \prod_{i=1}^n \left( 1-\frac{r_-}{\tau^i r_+} \right)  \notag\\ 
& \times
\int \cdots \int \prod_{i<j} \frac{z_i-z_j}{z_i-\tau z_j}
  \prod_{i=1}^n  \frac{F_{x,t}(z_i)}  {(1-\frac{z_i}{\tau\theta_+})(z_i-{\theta_-})} {\rm d}z_i
\label{ASEPmoment}
\end{align}
with $r_{\pm}$ defined below (\ref{cumgen}), $\th_{\pm} = \rho_\pm/(1-\rho_\pm)$ and 
\begin{equation}
 F_{x,t}(z) 
 =   \left( \frac{1+z}{1+z/\tau}\right)^x e^{-\frac{q(1-\tau)^2 z}{(1+z)(\tau+z)} t} .
\notag
\end{equation}
The contour of $z_i$ include  $-1,\tau \th_+$ and $\{ \tau z_j\}_{j>i}$ but 
  not  $-\tau, \theta_-$;
integrations are performed from  $z_n$ down to $z_1$, see Fig. 3.
This  contour   formula is a generalization of  the $\rho_-=0$ case  studied in \cite{BCS2014}.
See also a recent related work \cite{Aggarwal2016p}. 
The symmetric limit,  $\epsilon = 1-\tau \to 0$, is performed using the identity 
\begin{equation}
\sum_{j=0}^n (-1)^j \binom{n}{j} \langle \tau^{(n-j)N_\tau} \rangle
 =  \langle (1-\tau^{N_\tau})^n \rangle =
 \epsilon^n \langle N^n \rangle + o(\epsilon^n) 
\nonumber 
\end{equation} 
that  relates  the  $\tau$-moments  of   $N_{\tau}(x,t)$  in the ASEP
 to  the $n$-th moment  of the observable $N(x,t)$ in the SEP. 
 Each term  on the  left-hand side is   given by
 a  complex contour  integral, that  has to  be expanded
  with respect to  $\epsilon$.  This is  achieved first by   evaluating 
 the residues of the contour  integrals  at the poles in the vicinity of $\th_+$,
 leading to a formula in the form,
\begin{equation}
 \langle (1-\tau^{N_\tau})^n \rangle =\sum_{k=0}^n \mu_{n,k}(\epsilon)  J_k \epsilon^k
 \nonumber 
\end{equation} 
 where  the  combinatorial coefficients  $\mu_{n,k}(\epsilon)$ contain   the contributions  
 of the  residues and  the   $J_k$'s are  $k$-fold  integrals localized
 around the origin. Then, explicit  recursive relations
 for  the   $\mu_{n,k}(\epsilon)$'s  are found
  and   large time asymptotics  of the  $J_k$'s
 are  extracted. This allows us, finally, to obtain a formula for the $n$-th
  moment of  $N(x,t)$  and  for its
  $n$-th cumulant. The  expressions for the cumulants are given by 
\begin{align}
\label{cumulant}
 \frac{\langle N(x,t)^n \rangle_c}{\sqrt{t}}  \sim
 \sum_{l=1}^n  \frac{\alpha_{n,l}(r_+,r_-)}{\sqrt{l}} \Xi(-\sqrt{l}\xi) 
   -2\alpha_{n,l}(1,0) \xi \rho_+^l  
\end{align}
where $\Xi(\xi)$ was defined below (\ref{ldrBM}) and 
\begin{align}
 & \frac{ \alpha_{n,l}(a,b)}{(l-1)!}  =  (-1)^l 
\sum_{\substack{ \sum j l_j =n   \\   \sum  l_j = l} }
 \frac{n!}{ \prod_{j=1}^n l_j!} \,  \prod_{j=1}^n 
\left(\frac{a + (-1)^jb }{j!}\right)^{l_j}.
\label{def:alphanl}
\end{align} 
Taking the generating function of
 the cumulants leads to   \eqref{cumgen}. 
 The full details of the derivation will be given in \cite{InPreparation}. 

 \begin{figure}[t]
 \includegraphics[scale=0.7]{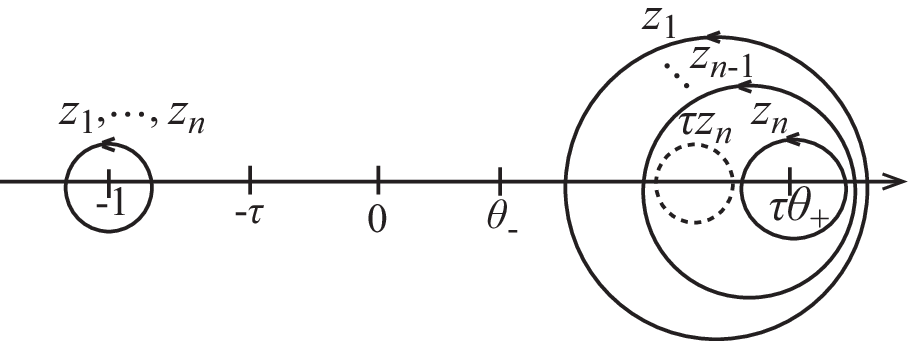}
 \caption{The integration contours in (\ref{ASEPmoment}). }
 \end{figure}

In this work, we  obtain   the  exact formula for the large deviations 
  of a tracer  in the  one dimensional 
symmetric simple exclusion process. This formula yields 
 all the cumulants of the tracer position, in the long time limit.
 This    answers  
  a  problem  that has  eluded solution for years
 \cite{Harris1965,Sethuraman2013}.
 Our results  are  valid both  when the system is  at equilibrium 
  with  uniform density,  and  when the system is
   out of equilibrium, starting with a step density profile,   the 
   tracer being  initially located at the boundary between the  two
 domains of unequal density. Some  of our  formulas  for  the  cumulants
 are prone  to  experimental tests,
 e.g. using  colloidal particles \cite{WBL2000}. They  can also
 be used  as benchmarks for numerical methods to
  evaluate large deviations, such as the one proposed in 
\cite{GiardinaKurchanPeliti2006}.  

 The  derivation of the central formula \eqref{cumgen} uses 
  the powerful
  mathematical arsenal of  integrable probabilities  developed
to solve the one-dimensional Kardar-Parisi-Zhang (KPZ)
  equation, the ASEP and related asymmetric models 
 \cite{SasamotoSpohnKPZ,Imamura2007, Imamura2012,CorwinRV,BCS2014,BC2014}.
Generalizations of the ideas and techniques in this article will allow us to 
reveal various intricate properties of the SEP and related symmetric models, 
which would have been difficult with other means. 

  Infinite  systems out of equilibrium  keep in general 
   the memory of 
  the initial conditions  \cite{Leibovich2013}.
  For the models in the KPZ universality class, it has been well established that 
  different  initial conditions  can lead to different statistical laws in the long time
 limit \cite{PS2002a, Sasamoto2008,CorwinRV,Takeuchi}. This
 must  also be true  in the tagged particle problem in the SEP and one would
 like to study more general set-ups  than the step   profile. 
 In particular, instead of taking  averages   over  an ensemble
 of  fluctuating initial step profiles and over the dynamics
  ({\it annealed case}), 
  one could start  with a  deterministic  
 initial configuration and average  only over the history of the process
  ({\it quenched  case}). For the latter case, even less is known 
  \cite{Gerschenfeld2009,KMS_tagged,DerridaSadhu} compared with 
  the former, but new progress is expected to be achieved 
  by extending our approach, combined with results for  
  the ASEP, e.g.  \cite{OQR2016}.  
 
Finally, we would like  to relate  our derivation to the
  macroscopic fluctuation  theory (MFT) \cite{Derrida2007,Bertini2014},
 one of the most promising  approaches to study
  systems far from equilibrium. The MFT  is based on a variational principle,
 that  determines  the optimal path that produces a given fluctuation, leading
 to two coupled  nonlinear 
 Euler-Lagrange equations.  For  reflecting  Brownian particles,
 these  equations can be  linearized  and solved, leading 
 to the large deviations of the tracer   \cite{KMS_tagged}.
  However, for the symmetric exclusion process, the MFT equations  are,
 for the moment, intractable.  Our exact calculations  
may give some hint to solve the MFT equations for this  non-linear case.

 \vskip 0.3cm 
The authors  are grateful  to   S. Mallick
 for a careful  reading of the manuscript
 and to  P. L. Krapivsky for interesting  discussions. 
 We thank the JSPS core-to-core program "Non-equilibrium 
dynamics of soft matter and information"  which initiated this work. 
Parts of this work were performed during  stays at ICTS  Bangalore and 
at KITP  Santa Barbara. 
This research was supported in part by the National Science Foundation under 
Grant No. NSF PHY11-25915. The works of T.I and T.S. are also supported by JSPS 
KAKENHI Grant Numbers JP25800215, JP16K05192 and P25103004, JP14510499, JP15K05203, JP16H06338 respectively.

%


\clearpage
\onecolumngrid
\begin{center}
\textbf{\large Supplemental note for  ``Large deviations of a tracer in
  the  symmetric exclusion process''}
\end{center}
\setcounter{equation}{0}
\setcounter{figure}{0}
\setcounter{table}{0}
\setcounter{page}{1}
\makeatletter
\renewcommand{\theequation}{S\arabic{equation}}
\renewcommand{\thefigure}{S\arabic{figure}}
\renewcommand{\bibnumfmt}[1]{[S#1]}
\renewcommand{\citenumfont}[1]{S#1}

 \begin{center}
 	This supplemental note contains  some extra-information 
  and  the derivation of some results given in the main text.
 \end{center}
 
\section{Large deviation function for non-interacting particles}

 When particles do not interact the motion of a tracer is a single-body
 problem.
 We recall the calculation  of  the large deviation function of a single
 particle  on a one-dimensional lattice,
 hopping  to the right and to the left with rates $p$ and $q$, respectively.
 We use the same notations as  in the main text.
 Discretizing  time   in infinitesimal steps $dt$, we
 write the position $X_t$ of the particle at time $t$ as
 $$ X_t = \epsilon_1 + \ldots  \epsilon_{t/dt}$$
 where the increments  $\epsilon_i$ are independent and can take
 three values, +1, -1 and 0 with probabilities $p  dt$, 
 $q  dt$ and $ 1 -(p+q)dt$, respectively. Because of  independence, we have
 $$ \langle e^{s X_t} \rangle  =   \langle e^{s \epsilon_i} \rangle^{t/dt}
  \sim e^{- t C(s)} $$
 with  $$C(s) = q( 1 -  e^{-s}) - p(e^{s} -1 ). $$
The Large Deviation Function, defined as
$$ \hbox{Prob}\left( \frac{X_t}{ t} = -\xi  \right)
  \simeq  \exp[-{t} \phi(\xi)]$$
is obtained from  $C(s)$  by  Legendre Transform and is given by
\begin{eqnarray}
 \phi(\xi) = p + q - \sqrt{\xi^2 + 4 p q} + 
 \xi \ln \frac{ \sqrt{\xi^2 + 4 p q} + \xi}{2 q}  \, .
\end{eqnarray}
 Note that the rate for large deviations is proportional to the time $t$
 as expected in a non-interacting system (whereas for the SEP we have
  a  $\sqrt{t}$  prefactor to the large deviations).

\section{Proof of the relation (9) between $X_t$ and $N(x,t)$}

We first define  $Q(x,t)$ the  time-integrated current
 that has flown through the bond $(x,x+1)$ between time 0 and $t$.
 The total current
  $Q(x,t)$ is equal to the total number of particles that have jumped
 from $x+1$ to $x$  {\it minus} the total number of particles that have jumped
 from $x$  to $x+1$ during the time interval $(0,t)$
 (because we take  the  convention that a current flowing from
right to left is counted positively). 

By particle conservation,
 we find that the  relation between   $Q(x,t)$ and  $N(x,t)$ is given by:
\begin{itemize}
\item for $x = 0$,  $Q(0,t) = N(0,t) = N_t$.
\item For $x > 0$, $Q(x,t) = Q(0,t) +  \sum_{y=1}^x \big(\eta_y(t) -\eta_y(0)\big)$, 
 or equivalently   $N(x,t) = Q(x,t) +  \sum_{y=1}^x \eta_y(0)$.
\item For $x < 0$, $N(x,t) = Q(x,t) -  \sum_{y=x+1}^0 \eta_y(0)$.
\end{itemize}
 Consider a site  $x>0$, located
 to the right of  $X_0$, the initial
 position of the tracer.
  For  the tracer $X_t$ to  be  to the right of $x$,
 it is necessary that all the particles that were initially  between  $X_0$  
 and $x$ have crossed the bond  $(x,x+1)$
 from left to right (including the tracer itself).
 This means that the  total current $Q(x,t)$ has to be less than
 $- \sum_{i=X_0}^x \eta_i(0) = - \sum_{i=1}^x \eta_i(0)$ [here  we use  the
 fact   the tracer is defined to be the particle 
  which at $t =0$ is  the closest to the origin from the right;
 therefore all sites between 1 and $X_0 -1$ are empty at $t =0$].
  We conclude that 
 for  $x \ge  X_0$, 
\begin{eqnarray}
 \hbox{Prob}\left( X_t > x \right) &=& 
 \hbox{Prob}\left(Q(x,t) \le - \sum_{i=1}^x \eta_i(0) \right)  \nonumber  \\
 &=&  \hbox{Prob}\left(N(x,t) \le 0 \right).
\label{eq:dte}
\end{eqnarray}
 A similar reasoning allows us to show that for  $x <  X_0$, 
\begin{eqnarray}
 \hbox{Prob}\left( X_t \le  x \right) 
=  \hbox{Prob}\left(N(x,t) >  0 \right).
\label{eq:gauche}
\end{eqnarray}
The two identities (\ref{eq:dte}-\ref{eq:gauche}) 
 imply   the relation (6) of  the main text.

\section{Relation between $\phi (\xi)$ and $\mu(\xi,\lambda)$}
Here we show that $\phi (\xi)$, the large deviation function of tracer\rq{}s 
position can be written in terms of $\mu(\xi,\lambda)$, 
the characteristic function of $N(x,t)$, as 
\begin{align}
\phi(\xi)=\max_{\lambda}\mu(\xi,\lambda).
\label{phimu}
\end{align}
For this purpose we introduce $\Phi(\xi,q)$,
the large deviation function of  $N(x,t)$, given by 
\begin{equation}
  \text{Prob}\left( \frac{N(x,t)}{\sqrt{t}} = q  \right)
  \simeq  \exp[-\sqrt{t} \Phi(\xi,q)] 
\label{def:PhiN}
\end{equation}
 with  $\xi = - \frac{x}{\sqrt{4 t}}.$ From~\eqref{eq:gauche} and~\eqref{def:PhiN}, we find
\begin{equation}
 \phi(\xi) = \,  \Phi(\xi,q =0). 
\label{phiPhiN}
\end{equation}
Note that the  two functions $\Phi$ and $\mu$ are Legendre transforms 
of each other 
\begin{equation}
   \Phi(\xi,q)  = \max_{\lambda}\left(\mu(\xi,\lambda) +  \lambda q  \right).
\label{PhiVERSUSmu}
\end{equation}
Thus from~\eqref{phiPhiN} and~\eqref{PhiVERSUSmu}, we get~\eqref{phimu}.

\section{Proof of the fluctuation relation}

 We start with the parametric representation of the large deviation
 function (eq. (5) in the main text). The function $\phi(\xi)$, for a given
 value of $\xi$ is  given
 by $  \phi(\xi)=  \mu(\xi, \lambda^*) $  where 
   $\lambda^*(\xi)$ is 
 such that $ \frac{ \partial \mu(\xi, \lambda^*)} {\partial  \lambda}  =  0$,
 which, using  the formula (6) for $\mu(\xi, \lambda)$, is equivalent to
\begin{equation}
\left( r_+ e^{\lambda^*} - r_{-} e^{-\lambda^*} \right) 
 \sum_{n=1}^{\infty}  \frac{ (-1)^{n -1}\omega(\lambda^*)^{n -1}}{n^{1/2}}  A(\sqrt{n}\, \xi) \,
 =  \xi \frac{ \omega(\lambda^*) + \rho_+  + \rho_{-}  } {\omega(\lambda^*) + 1}. 
\label{derivMU}
\end{equation}
 First, we note under the change $\lambda\rightarrow
\log \frac{r_{-}}{r_{+}}\ - \lambda$,
\begin{equation}
 \omega\Big(\log \frac{r_{-}}{r_{+}}\ - \lambda\Big) = \omega(\lambda),~~r_+ e^{\log \frac{r_{-}}{r_{+}} - \lambda} - r_{-} e^{-\log \frac{r_{-}}{r_{+}} + \lambda}
 =  - \left( r_+ e^{\lambda} - r_{-} e^{-\lambda} \right). 
\end{equation}
Using these properties  and $A(\xi)=A(-\xi)$ to~\eqref{derivMU},
we deduce that 
\begin{equation}
\lambda^*(-\xi)  = \log \frac{r_{-}}{r_{+}}\ - \lambda^*(\xi).
\end{equation}
 Therefore we have
$$  \phi(\xi) - \phi(-\xi) = \mu(\xi, \lambda^*(\xi)) -
   \mu(-\xi, \lambda^*(-\xi)) = 
 \mu(\xi, \lambda^*(\xi)) - 
  \mu\Big(-\xi, \log \frac{r_{-}}{r_{+}}\ - \lambda^*(\xi)\Big). $$
Using again the formula (6) and the fact that $\omega$ is invariant,
 we find that this expression is equal to
 $$  \xi \log\frac{\left(1+\rho_+(e^{\lambda}-1) \right) 
 \left(1+\rho_+(e^{\lambda'}-1) \right)  } 
{\left(1+\rho_-(e^{-\lambda}-1)\right)\left(1+\rho_-(e^{-\lambda'}-1)\right) },$$  
where we have written $\lambda$ for  $\lambda^*(\xi)$ and
 $\lambda'$ for  $\lambda^*(-\xi)$. After simplification this expression
 reduces to 
$$2 \xi \log\frac{ 1 - \rho_{+}}{1 - \rho_{-}} $$
 thus proving the Fluctuation Theorem.

\section{Calculation of the higher cumulants of  $X_t$ } 
To extract the cumulants of the $X_t$ from $\mu(\xi,\lambda)$, 
it is useful to use a parametric representation
for the cumulant  generating function $C(s)$: 
\begin{eqnarray}
  \frac{ \partial \mu(\xi, \lambda)} {\partial  \lambda}   &  =  & 0,  
\label{premeq} \\
   \frac{ \partial \mu(\xi, \lambda)} {\partial  \xi}  &  = &  2s,
\label{deuxeq}  \\
    C(s)  & = & 2s \xi  +  \mu(\xi, \lambda).   \label{troiseq}
\end{eqnarray}
 The first two equations define implicitly two functions
 $ \xi(s)$  and $\lambda(s)$, that, after substitution in the third
 equation, provide us with the cumulant  generating function $C(s)$.

\subsection{Equilibrium case with  uniform density}

 For a stationary initial condition,  $\rho_{+} =\rho_{-} = \rho $,
 the functions $\xi(s)$ and $\lambda(s)$  vanish
 when $s \to 0$. The  strategy is to write power-series
 expansions w.r.t. $s$ for these two functions
\begin{eqnarray}
 \xi(s) &  =  &  \sum_{i \ge 1} x_i s^i,
 \nonumber \\ 
 \lambda(s)   &  =  & \sum_{i \ge 1} l_i s^i. \nonumber
\end{eqnarray}
 In order to calculate the unknown coefficients $x_i$'s and $l_i$'s, 
 we substitute these expansions in \eqref{premeq} and in \eqref{deuxeq}.
 At  each order $n$,  we obtain  two inhomogeneous  linear  equations 
 for  $x_n$ and $l_n$, the r.h.s 
 of which  involve higher powers of  $x_i$'s and $l_i$'s
 with $i < n$. This system can be solved systematically to any desired
 order. This can be done by hand up to $n=4$ and for higher orders
 by using a symbolic 
 computation software (such as Mathematica).
 Substituting the series truncated  at order $n$ 
 for  $\xi(s)$ and $\lambda(s)$ in  $C(s)$
 using \eqref{troiseq} gives  the values
 of the first $n$ cumulants. 
All  odd cumulants vanish and the first few even cumulants are given by
the following expression.
\hfill\break
At order 2:
$$  \frac{\langle X_t^2 \rangle_c}{\sqrt{4 t}} = 
   \frac{1-\rho}{\rho \sqrt{\pi} } .  $$ 
\hfill\break
At order 4:
\begin{equation}
 \frac{\langle X_t^4 \rangle_c}{\sqrt{4 t}} = 
  \frac{1-\rho}{\sqrt{\pi} \rho^3}
   [1-(4-(8-3\sqrt{2}) \rho)
  (1-  \rho)+\dfrac{12}{\pi}(1-\rho)^2]. \nonumber 
\end{equation}
\hfill\break
At order 6:

\begin{eqnarray}
 \frac{\langle X_t^6 \rangle_c}{\sqrt{4 t}} = 
 \frac{1 -\rho}{ \pi^{5/2} \rho^5} \Big[
&&  \left(1020 - 450 \pi + 45\pi^2  \right)  \nonumber \\
  -  \rho && \left(4800   -  \pi(2700   - 540  \sqrt{2})
 + \pi^2 (270  - 45   \sqrt{2}) \right) 
 \nonumber \\
   +  \rho^2 &&\left(6120 -  \pi(5250  - 1620 \sqrt{2})
 +  \pi^2 (570    - 225  \sqrt{2} + 40   \sqrt{3}) \right)
 \nonumber \\
 -  \rho^3 && \left(4080 - \pi (4200  - 1620 \sqrt{2})
+ \pi^2(480  - 300 \sqrt{2} + 80  \sqrt{3}) \right)
 \nonumber \\
  +  \rho^4 && \left( 1020 - \pi( 1200  - 540 \sqrt{2})
+ \pi^2( 136  - 120   \sqrt{2} + 40   \sqrt{3}) \right)
\Big]. \nonumber
\end{eqnarray}

\subsection{Non-equilibrium case with $\rho_{+} \neq \rho_{-}$ }

   When the system starts with an  initial step-profile, it remains
 out-of-equilibrium  and the tracer drifts away from the origin
  as shown  by (12) in the main text.  In order to calculate
 the cumulants we must again solve the system \eqref{premeq},
 \eqref{deuxeq} and \eqref{troiseq}. Here,  when $s \to 0$,  we still
 have $\lambda \to 0$ but $\xi \to \xi_0$ where  $\xi_0$ satisfies
(13) in the main text.
  The procedure explained above for the equilibrium case
 can still be applied but if one just wants to calculate the variance of 
 the tracer's position it is simpler  
 to use the parametric representation of the large deviation
 function:
 $ \phi(\xi)=  \mu(\xi, \lambda^*) $  where 
   $\lambda^*(\xi)$ is 
 such that $ \frac{ \partial \mu(\xi, \lambda^*)} {\partial  \lambda}  =  0$
  (eq. (8) in the main text).

 The large deviation function  is strictly positive
 for $\xi \neq \xi_0$ and 
 vanishes at  $\xi_0$, i.e.,
 $ \mu(\xi_0, \lambda^*(\xi_0))  = 0 $. It is elementary to check
 from the formula  (6) in the main text that 
  $\lambda^*(\xi_0) = 0$.

 To calculate  the variance of $X_t$, we perform a second
 order expansion    of $ \phi(\xi_0 + \epsilon)$
 with respect to  $\epsilon$.
 Writing $\lambda^*(\xi_0 + \epsilon) \simeq 1 + a  \epsilon$,
  we determine $a$  such that 
 $  \partial_\lambda \mu(\xi_0 + \epsilon, \lambda^*(\xi_0 + \epsilon)))$
  vanishes  at first  order   in  $\epsilon$. We find that  $a = -U/V$ with 
 $$ U = \rho_{+} \erfc{(\xi_0)} + \rho_{-} \erfc{(-\xi_0)} $$
and
 $$ V= 
 \frac{\rho_{+}^3 +  \rho_{-}^3 - 3 \rho_{+}^2 \rho_{-} -3 \rho_{+} \rho_{-}^2
 + 4 \rho_{+} \rho_{-}}  { (\rho_{+} +\rho_{-})}   A(\xi_0) 
   -  (\rho_{+} -\rho_{-})^2  \frac{A(\sqrt{2} \, \xi_0)}{\sqrt{2}}. $$
 Substituting this result in  $\phi(\xi_0 + \epsilon) \simeq
 \mu(\xi_0 + \epsilon, 1 + a  \epsilon)$  gives the correct  dominant 
 term in the large deviation function at order $\epsilon^2$:
  $$\phi(\xi) \simeq \frac{U^2}{2 V} (\xi -\xi_0)^2 \, .$$
 From this Gaussian limiting form, we deduce that
 $$\mathrm{Var} (\xi) =  \frac{V} {U^2 \sqrt{t}} \, $$
 which, after rearranging the terms,  leads to the formula
 for the variance  of the tracer,
 given in the main text:  
$$ \mathrm{Var} (X_t) =  \frac{ 4 K (\rho_{+}  - \rho_{-})^2  A(\xi_0) \sqrt{ t}}
{(\rho_{+} \erfc{(\xi_0)} + \rho_{-} \erfc{(-\xi_0)})^2} $$ 
 with 
$$K =
 \frac{\rho_{+}^3 +  \rho_{-}^3 - 3 \rho_{+}^2 \rho_{-} -3 \rho_{+} \rho_{-}^2
 + 4 \rho_{+} \rho_{-}}  { (\rho_{+} +\rho_{-}) (\rho_{+}  - \rho_{-})^2}   
-  \frac{A(\sqrt{2} \, \xi_0)}{\sqrt{2} A(\xi_0)}   \, . 
$$

%
%

\end{document}